\documentstyle
[psfig,epsf]
{l-aa}
\begin{document}

\def\th{$^{13}$}
\def\ei{$^{18}$}
\def\tw{$^{12}$}
\def\Lcs{{\hbox {$L_{\rm CS}$}}}
\def\lunits{K\thinspace \kms\thinspace pc$^2$}

\def\,{\thinspace}
\def\etal{et al.}


\def\kms{km\thinspace s$^{-1}$}
\def\Lsun{L$_\odot$}
\def\Msun{M$_\odot$}
\def\ms{m\thinspace s$^{-1}$}
\def\percc{cm$^{-3}$}

\font\sc=cmr10

\def\CBR{{\rm\sc CBR}}
\def\FWHM{{\rm\sc FWHM}}
\def\HI{{\hbox {H\,{\sc I}}}}
\def\HII{{\hbox {H\,{\sc II}}}}


\def\Ha{H$\alpha$}                      
\def\Htwo{{\hbox {H$_2$}}}
\def\nHtwo{n(\Htwo)}
\def\He#1{$^#1$He}                      
\def\water{H$_2$O}
\def\flecha{\rightarrow}
\def\12CO{$^{12}$CO}
\def\COJ#1#2{{\hbox {CO($J\!\!=\!#1\!\rightarrow\!#2$)}}}
\def\CO#1#2{{\hbox {CO($#1\!\rightarrow\!#2$)}}}
\def\CeiO#1#2{{\hbox {C$^{18}$O($#1\!\rightarrow\!#2$)}}}
\def\CSJ#1#2{{\hbox {CS($J\!\!=\!#1\!\rightarrow\!#2$)}}}
\def\CS#1#2{{\hbox {CS($#1\!\rightarrow\!#2$)}}}
\def\HCNJ#1#2{{\hbox {HCN($J\!\!=\!#1\!\rightarrow\!#2$)}}}
\def\HCN#1#2{{\hbox {HCN($#1\!\rightarrow\!#2$)}}}
\def\HNC#1#2{{\hbox {HNC($#1\!\rightarrow\!#2$)}}}
\def\HNCO#1#2{{\hbox {HNCO($#1\!\rightarrow\!#2$)}}}
\def\HCOpJ#1#2{{\hbox {HCO$^+$($J\!\!=\!#1\!\rightarrow\!#2$)}}}
\def\HCOp#1#2{{\hbox {HCO$^+$($#1\!\rightarrow\!#2$)}}}
\def\HCOpp{{\hbox {HCO$^+$}}}
\def\J#1#2{{\hbox {$J\!\!=\!#1\rightarrow\!#2$}}}
\def\noJ#1#2{{\hbox {$#1\!\rightarrow\!#2$}}}


\def\Lco{{\hbox {$L_{\rm CO}$}}}
\def\Lhcn{{\hbox {$L_{\rm HCN}$}}}
\def\Lfir{{\hbox {$L_{\rm FIR}$}}}
\def\Ico{{\hbox {$I_{\rm CO}$}}}
\def\Sco{{\hbox {$S_{\rm CO}$}}}
\def\Ihcn{{\hbox {$I_{\rm HCN}$}}}


\def\Tastar{{\hbox {$T^*_a$}}}
\def\Tmb{{\hbox {$T_{\rm mb}$}}}
\def\Tb{{\hbox {$T_{\rm b}$}}}

\title{On the origin of the high velocity SiO maser emission from late-type
stars}

\author {
F. Herpin\inst{1},
A. Baudry\inst{1},
J. Alcolea\inst{2},
J. Cernicharo\inst{3}
}

\offprints{F. Herpin, herpin@observ.u-bordeaux.fr}

\institute {Observatoire de Bordeaux, BP 89, F-33270 Floirac, France
\and
Observatorio Astron\'omico Nacional, Apartado 1143, E-28800 Alcal\'a de
Henares, Spain
\and
CSIC, IEM, Depto. F\'{\i}sica Molecular, Serrano123, E-28006 Madrid, Spain}

\thesaurus{07 (02.12.1, 02.12.3, 02.13.3, 08.03.4, 08.12.1)}

\date{Received 28 October 1997; accepted 3 March 1998}

\maketitle
\markboth{High velocity SiO maser emission}{High velocity SiO maser emission}
\begin{abstract}

We have undertaken toward 30
Mira or semi-regular variables and one OH/IR object
highly sensitive observations of the $v=1, J=2
\rightarrow 1$ and $3 \rightarrow 2$ transitions of SiO
simultaneously with observations of the $J = 1 \rightarrow 0$ transition
of CO during three observing sessions in the period 1995 to 1996. As
in our previous observations of 1994, we observe that
for several stars the SiO profiles exhibit unusually broad wings
which sometimes exceed the terminal velocity of the expanding
circumstellar envelope
traced by the thermal CO emission. We have discovered a clear dependence
of the SiO wing emission on the
optical phase. These wings are probably due to peculiar gas motions
and varying physical conditions in relation with the stellar pulsation.
However, we cannot exclude
other mechanisms contributing to the observed wings. In particular,
SiO turbulent motions for the semi-regular variables or the
asymmetric mass loss mechanism may play a role. We conclude that the
SiO wing emission is due to masing processes and that this emission
very likely arises from the inner part of the circumstellar envelope.


\end {abstract}

\section{Introduction}

 The SiO molecule exhibits widespread maser emission
from O-rich circumstellar envelopes (CEs) around Long Period Variables (LPVs).
The variety of rotational transitions emitted from several vibrational
levels and the strength of the emission  make these
lines very useful for a study of the innermost layers of CEs. Recently,
during sensitive SiO
observations of several O-rich late-type stars, we discovered
unexpectedly broad wing SiO
emission (Cernicharo et al. 1997, hereafter referred as CABG). We found
that the
$v=1, J=2 \rightarrow 1$ SiO emission reaches and sometimes exceeds the
maximum velocity traced by the quasi-thermal emission of the CO molecule.
Several mechanisms may be invoked to
explain such wings: turbulent motions, rotation of dense SiO clumps, high
velocity shocks produced during the pulsation of the star, or high
velocity bipolar ejection of gas from the star. These mechanisms were
discussed by CABG, but no firm conclusions were reached, although it
was recognized that the
pulsation of the star and asymmetric mass loss could play an important
role. Obviously interferometric
observations and a monitoring of the SiO wing emission are needed to
shed light on the location and the physical origin of the high velocity SiO
emission.

The main purpose of this paper is to present and analyze new data
gathered at different epochs on the SiO velocity wings in order to
study the physical mechanisms responsible for this phenomenon, and, at the
same
time, to obtain deeper insight
into the complex kinematics of the circumstellar shells around
late-type stars. The latter question is clearly important since it is
related to other
crucial problems such as the expansion of the CE and the mass loss
\begin{figure*} [ht]
  \begin{center}
     \epsfxsize=15.cm
     \epsfbox{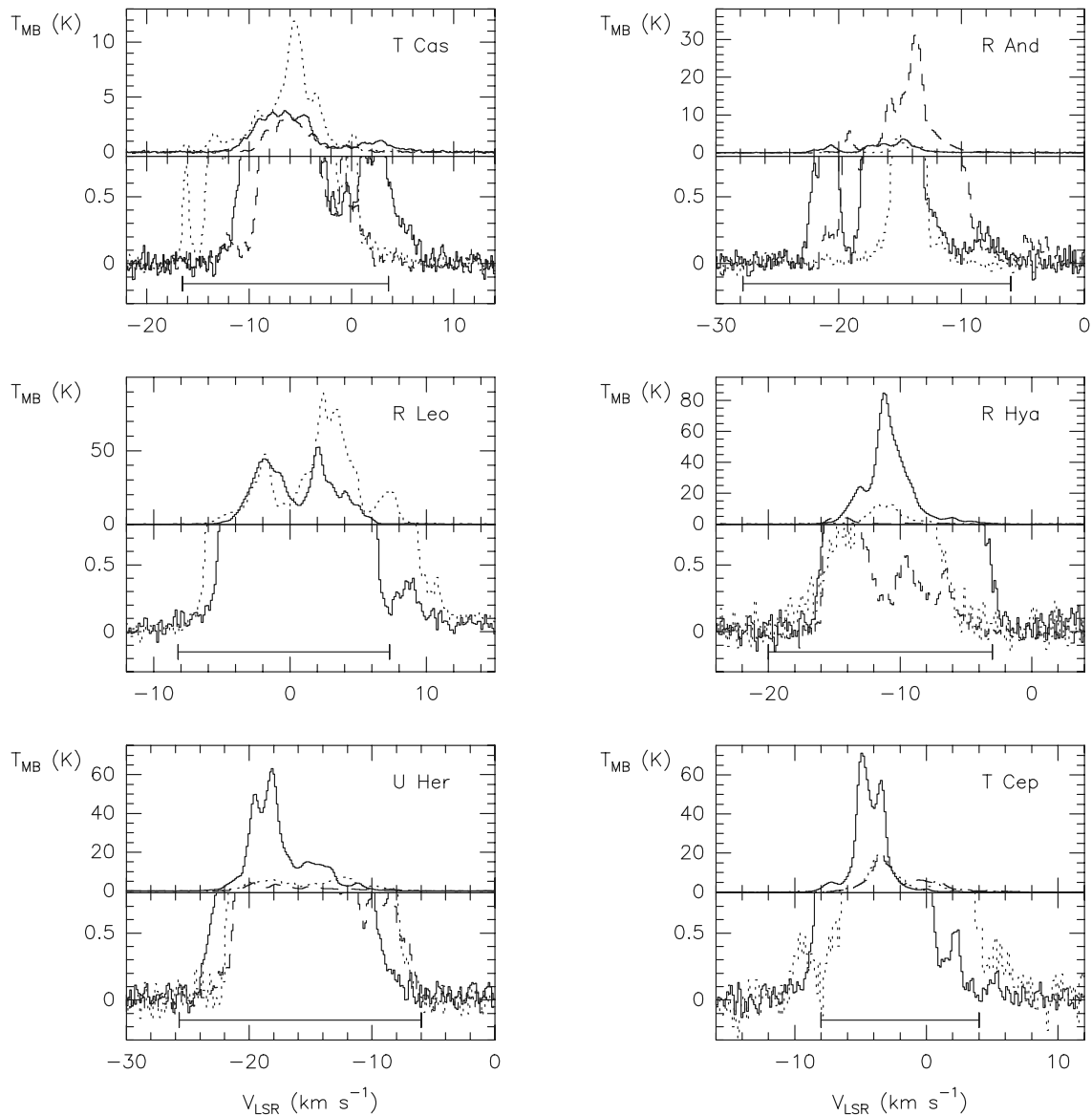}
  \end{center}
  \caption []{Examples of SiO $v=1, J=2 \rightarrow 1$ spectra for
  {\em Mira variables} taken with $0.136$ {\kms} spectral resolution in
June 1995
(continuous line), April 1996 (dashed line) and October 1996
(dotted line). For each star
the top panel shows the full main beam brightness temperature scale
and the bottom panel corresponds to an enlargement of the same data,
with the CO line emission width ($\Delta V(CO)$ above the $2\sigma$ level)
represented by a horizontal segment.
For R Leo, no data are available for April 1996. To transform the
intensities into Jy one must multiply by $4.4$ Jy/K.}
  \label{asio_spectra}
\end{figure*}
phenomenon, or the processes leading to the formation of dust.
We have considered a rather small but homogeneous sample of stars as it
includes Miras and a few semi-regulars for which all stellar
characteristics (mass loss, temperature, spatial distribution \ldots)
are uniformly represented. We have monitored the SiO maser line
profiles
in order to investigate the
relation of broad (weak) emission in wings with the stellar light
phase. In particular, we wish to test whether the shocks driven by the
stellar pulsation could play a dominant role in the
occurrence of SiO high
velocity features. In Sects. 2 and 3 we give details of our observations and
present our main results. In Sect. 4 we discuss the
different hypotheses which could explain the SiO line wings and their
location.
\newpage

\section{Observations}

\begin{figure*} [t]
  \begin{center}
     \epsfxsize=15cm
     \epsfbox{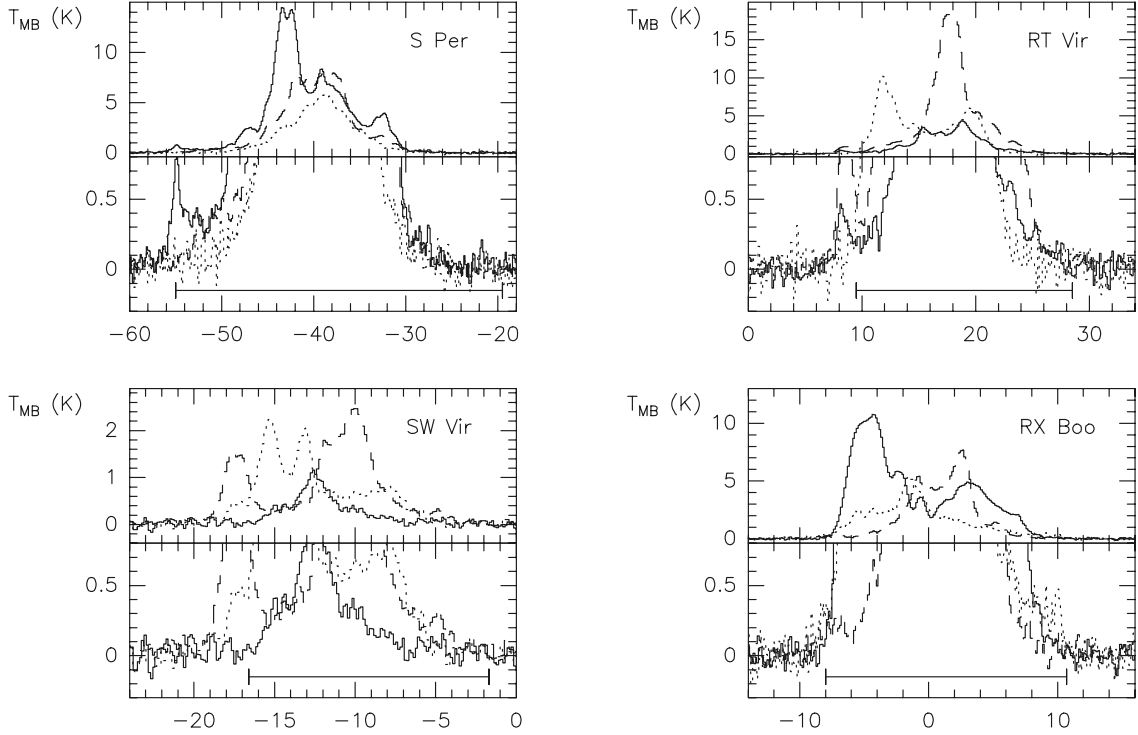}
  \end{center}
  \caption []{Examples of SiO $v=1, J=2 \rightarrow 1$ spectra for
  {\em Semi-Regular variables} taken with $0.136$ {\kms} spectral
resolution in June 1995
(continuous line), April 1996 (dashed line) and October 1996
(dotted line). For each star
the top panel shows the full main beam brightness temperature scale
and the bottom panel corresponds to an enlargement of the same data,
with the CO line emission width ($\Delta V(CO)$ above the $2\sigma$ level)
represented by a horizontal segment. To transform the
intensities into Jy one must multiply by $4.4$ Jy/K.}
  \label{bsio_spectra}
\end{figure*}
Our observations were carried out with the IRAM 30-m telescope in
June 1995 and in April and October 1996. They were combined with our earlier
observations of January 1994 first presented by CABG. The sample
of selected Mira and semi-regular variable stars was
identical for all epochs, apart from a few exceptions. The supergiant VY CMa
was not observed in 1995 and 1996.
In June 1995 and later, we added the Mira variable
$\chi$ Cyg to our sample. In this work, we have observed just one OH/IR object,
OH$127.8-0.0$, at one epoch only, June 1995. The sample of sources discussed
here is listed in the first column of Table 1. Two SIS receivers were used
simultaneously
at 86.243 and 230.538 GHz to observe the SiO $v=1, J=2
\rightarrow 1$ and the CO $J=2 \rightarrow 1$ lines. In 1995 and 1996 we
added a third receiver to observe the SiO $ v=1, J=3 \rightarrow 2$ line
at 129.363 GHz.
To properly detect and analyze weak SiO line wing signals we adjusted the
side-band noise level of the receiver phase-lock loops below $-$35 dB.
The 3, 2 and 1 mm receivers were SSB tuned and the attenuations of the
image band were of order 35 and 7 dB for the 3 and 2 mm SiO lines,
and 30 dB for the CO line. The best observing period was that of January
1994 (CABG) with low ambient
temperature (nearly $-10^\circ$ C) and almost no humidity. For the three
other epochs
the observations were made with clear sky but with relative humidity in
the range $30-70\%$. Depending on the source elevation and on the epoch,
the SSB system temperatures were in the range
$140-400$, $270-600$ and $180-1000$ K for the 3, 2 and 1 mm receivers,
respectively. Line calibration was deduced from regular observations of a
hot and cold load and from measurements of the sky emissivity. In this work
the line intensities are calibrated in terms
of the main beam brightness temperature. For the observations of 1995 and
1996 we
adopted 0.75 and 0.58 for the
telescope main beam efficiencies at the frequencies of the SiO $v=1, J=2
\rightarrow 1$ and $J=3 \rightarrow 2$ lines, respectively. In 1994 we used
0.60 for the
3 mm observations (CABG). These intensities (in K) can be transformed
into flux densities (in Jy) by multiplying by 4.4 at 3 and 2 mm. In
general, the pointing corrections were determined by
cross scanning the 3 mm maser line itself, using several filter
channels as a continuum detector. The
absolute pointing of the 3 mm receiver and the alignment between all
receivers lay in the range $2-4"$. The focus adjustment was regularly
monitored.
%
\scriptsize
\begin{table*} [t]
 \hspace* {-0.1 cm}{{\scriptsize{
  \caption{Selected stars and observed line parameters for the SiO $v=1, J=2
  \rightarrow 1$ and $v=1, J=3\rightarrow 2$ emissions observed in
  June 1995. Stars are arranged in ascending right ascension order. To
transform the
peak line intensities into Jy one must multiply the observed $J=2-1$
and $J=3-2$ values of $T_{MB}$ by $4.4$ Jy/K.}
 \begin{tabular}{lcccccc|ccccc} \hline \hline
 \multicolumn{7}{c}{\small {\bf SiO v=1 J=2-1}} &
\multicolumn{5}{c}{\small {\bf SiO v=1 J=3-2}} \\
\multicolumn{7}{c}{\small {\bf (June 1995)}} & \multicolumn{5}{c}{\small
{\bf (June 1995)}}
\\ \cline{3-12}
{\footnotesize  {\bf Stars}} &{\footnotesize  {\bf Stellar}} & {\footnotesize
{\bf $\sigma$}} &
{\footnotesize  {\bf $T_{MB}$}} & {\footnotesize {\bf $V_{LSR}$}}
&{\footnotesize
{\bf $\Delta V(2 \sigma)$}} & {\footnotesize  {\bf $\int T_{MB} dV$}} &
{\footnotesize  {\bf $\sigma$}} &
{\footnotesize  {\bf $T_{MB}$}} & {\footnotesize  {\bf $V_{LSR}$}} &
{\footnotesize {\bf $\Delta V(2 \sigma)$}} & {\footnotesize  {\bf $\int
T_{MB} dV$}} \\
   & {\footnotesize  {\bf Class}} & {\footnotesize  mK} &
   {\footnotesize K} &
   {\footnotesize kms$^{-1}$} & {\footnotesize kms$^{-1}$} &
   {\footnotesize Kkms$^{-1}$} &
   {\footnotesize mK} & {\footnotesize K} &
   {\footnotesize kms$^{-1}$} & {\footnotesize kms$^{-1}$} &
   {\footnotesize Kkms$^{-1}$} \\ \hline
{\bf Y Cas} & Mira & 34 & 7.3 & -19.1 & 15.3 & 26 & 51 & 3.9 & -18.8 &
10.4 & 12 \\
{\bf T Cas} & Mira & 33 & 3.7 & -6.4 & 19.8 & 27 & 53 & 1.3 & 2.7 &
14.3 & 5 \\
{\bf R And} & Mira & 32 & 3.4 & -14.8 & 18.5 & 15 & 49 & 2.1 &
-17.2 & 14.9 & 6 \\
{\bf IRC+10011} & Mira & 38 & 11.6 & 6.7 & 9.7 & 30 & 57 & 5.1 &
7.9 & 8.1 & 11 \\
{\bf $o$ Ceti} & Mira & 25 & 82.9 & 43.3 & 14.2 & 420 & 41 &
26.1 & 48.1 & 9.8 & 59  \\
{\bf S Per} & SRc & 39 & 14.5 & -43.2 & 32.0 & 107 & 56 &
4.3 & -32.0 & 25.8 & 36 \\
{\bf NML Tau} & Mira & 59 & 87.9 & 35.7 & 13.2 & 343 & 114 &
46.3 & 36.2 & 9.3 & 148 \\
{\bf TX Cam} & Mira & 85 & 38.3 & 10.2 & 12.4 & 118 & 51 &
15.1 & 11.7 & 10.7 & 38 \\
{\bf U Ori} & Mira & 38 & 2.3 & -36.8 & 13 & 13 & 72 &
4.1 & -44.6 & 9.8 & 6 \\
{\bf V Cam} & Mira & 44 & 6.4 & 6.5 & 18.3 & 22 &
82 & 2.4 & 7.2 & 10.1 & 6 \\
{\bf GX Mon} & Mira & 41 & 9.7 & -10.0 & 13.0 & 36 & 73 &
5.0 & -9.1 & 12.3 & 16 \\
{\bf S CMi} & Mira & 31 & 2.3 & 50.1 & 11.5 & 8 & 92 & 0.4 &
54.4 & 4.5 & 1 \\
{\bf W Cnc} & Mira & 35 & 1.3 & 34.8 & 10.8 & 6 & 87 & 0.6 &
35.7 & 2.8 & 1 \\
{\bf R LMi} & Mira & 30 & 10.7 & -3.8 & 15.8 & 48 & 51 & 2.4 &
1.1 & 10.9 & 11 \\
{\bf R Leo} & Mira & 52 & 50.2 & 2.0 & 15.8 & 260 & 52 &
19.4 & -0.8 & 13.7 & 69 \\
{\bf R Crt} & SRb & 36 & 5.1 & 8.7 & 25.1 & 39 & 65 & 1.0 &
6.1 & 16.0 & 6 \\
{\bf RT Vir} & SRb & 30 & 4.3 & 18.9 & 20.0 & 25 & 56 & 0.9 &
15.8 & 10.9 & 4 \\
{\bf SW Vir} & SRb & 35 & 1.1 &  -12.4 & 11.2 & 4 &    &      &
&      &     \\
{\bf R Hya} & Mira & 46 & 83.6 & -11.2 & 15.3 & 237 & 111 &
27.1 & -10.7 & 14.3 & 90 \\
{\bf W Hya} & SRa & 42 & 64.4 & 41.7 & 16.7 & 382 & 99 & 72.9 &
42.3 & 18.2 & 229 \\
{\bf RX Boo} & SRb & 35 & 10.8 & -4.4 & 18.2 & 72 & 55 & 0.5 &
1.2 & 7.9 & 1 \\
{\bf S CrB} & Mira & 54 & 7.0 & 0.3 & 11.5 & 16 & 65 & 0.2 & -1.0
& 5.1 & 1 \\
{\bf RU Her} & Mira & 33 & 3.9 & -10.6 & 15.1 & 21 & 66 & 2.3 &
-17.8 & 14.0 & 8 \\
{\bf U Her} & Mira & 45 & 61.3 & -18.1 & 16.4 & 221 & 50 &
8.1 & -18.5 & 14.3 & 58 \\
{\bf VX Sgr} & SRc & 47 & 30.3 & 7.2 & 35.9 & 263 & 74 & 15 &
0.0 & 33.1 & 139 \\
{\bf R Aql} & Mira & 33 & 2.9 & 50.3 & 16.1 & 18 & 62 & 0.5 &
44.4 & 8.1 & 2 \\
{\bf $\chi$ Cyg} & Mira & 55 & 29.0 & 11.1 & 16.4 & 93 & 53 &
19.3 & 11.8 & 18.2 & 73 \\
{\bf T Cep} & Mira & 36 & 67.2 & -5.1 & 16.1 & 167 & 55 &
6.4 & -3.3 & 12.1 & 17  \\
{\bf $\mu$ Cep} & SRc & 39 & 8.5 & 24.1 & 18.0 & 56 & 52 &
15.6 & 25.0 & 18.2 & 62 \\
{\bf R Cas} & Mira & 35 & 40.3 & 25.4 & 15.0 & 126 & 54 & 4.8
& 30.0 & 14.9 & 22 \\
{\bf OH127.8-0.0} & OH/IR & 31 & 0.4 & -56.5 & 8.3 & 1 & 15 &
0.1 & -58.3 & 10.2 & 0.6 \\ \hline \hline
 \end{tabular}}}}
 \label{table95}
\end{table*}
\normalsize
 For spectral analysis we used analog filters with both low (1 MHz) and
high (100 kHz)
resolution and the 2048-channel autocorrelator. The resolution thus
achieved was 3.5, 0.35, 0.136 and 0.07 {\kms} in SiO ($J=2-1$), 0.18 and
2.3 {\kms} in SiO ($J=3-2$) and 1.3, 0.4, 0.1 and 0.05 {\kms} in
CO($J=2-1$). We checked that by smoothing the autocorrelator channels to
a resolution similar to that given by
the filterbank channels the linewidths were identical. Because we searched
for line wing emissions as weak as a few tens of mK flat spectral
baselines are essential. All observations were thus made with the
wobbling secondary mirror system.
Figures \ref {asio_spectra} and \ref {bsio_spectra} shows examples
of SiO $v=1, J=2 \rightarrow 1$ spectra
obtained in 1995 and 1996. Note that the quality
of the spectral baselines facilitates the comparison of the CO and SiO
line wings.

\section{Results}
\subsection{SiO $v=1, J=2 \rightarrow 1$ emission}
 In Table 1 we present the SiO $v=1, J=2 \rightarrow 1$ line
parameters of sources observed in June 1995 (including OH$127.8-0.0$ at the
bottom of
Table 1). The line parameters for April and October 1996 are given in
Herpin (Ph.D. Thesis 1998). The SiO and CO line parameters of 1994 are
given in Table 1 of CABG. In column 2 of Table 2 of the present work we
also list the line
width above the $2\sigma$ noise level of the CO observations made in 1994
because they are used
as our CO reference data. Although the atmosphere
was less transparent at the CO($2-1$) line in 1995 and 1996 than in 1994
we verified that the CO central velocity and linewidth remained constant with
the epoch of the observations within a few $\%$.
The SiO main beam brightness temperature $T_{\rm MB}$ and velocity $V_{\rm
LSR}$ given in Table 1 correspond to the strongest emission
feature in the spectra. The SiO full linewidth above the $2\sigma$ spectral
noise level
$\Delta V$ is also given in Table 1 for June 1995. As in
CABG, our new observations show that several stars exhibit pronounced
blue or red wings. This is a firm result because it is confirmed by the
different backends and spectral resolutions available on the 30-m. These high
velocity wings (e.g. Figs. \ref{asio_spectra} and \ref {bsio_spectra}) are
weak compared to
the bulk of the SiO $v=1, J=2 \rightarrow 1$
emission whose central velocity is close to the CO central velocity;
the latter gives the systemic velocity of the underlying star. We have no
direct
proof at the moment that the SiO line wing emission is masing as are the
main velocity features. We note that $v=1$ thermal lines would be very
difficult to
detect at these high velocities where no thermal
SiO emission is observed in the ground vibrational state (Cernicharo
\etal 1994, Bujarrabal \etal 1989). Furthermore, in the case of R Leo
(CABG) and
some other stars, the linewings are polarized. This indicates maser
amplification, a fact strengthened by the line variability of the
line wing emission discussed in Sect. \ref{discussion}.
\begin{figure*} [t]
  \begin{center}
     \epsfxsize=16cm
     \epsfbox{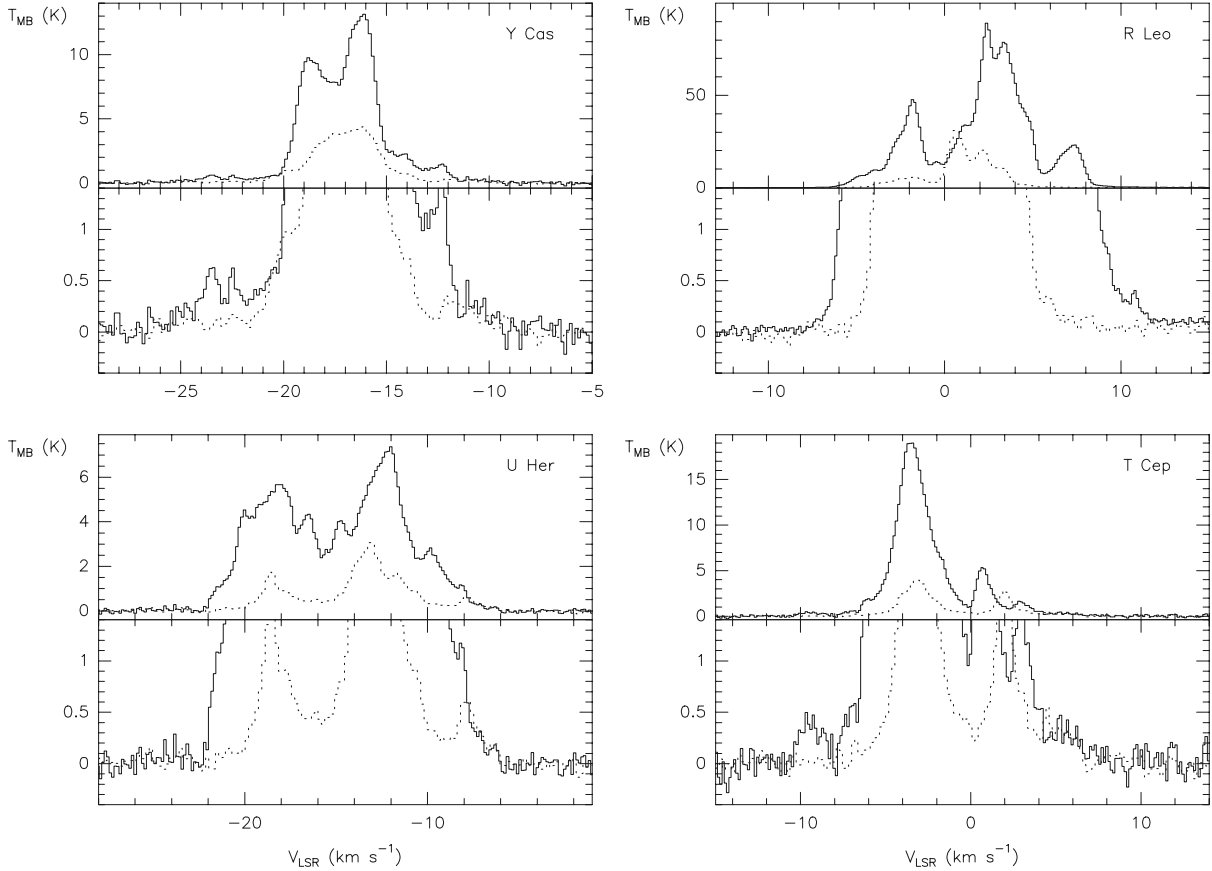}
  \end{center}
  \caption []{SiO $v=1, J=3 \rightarrow 2$ spectra (dashed line)
observed toward Y Cas, U
Her, R Leo and T Cep superposed on the $J=2 \rightarrow 1$ spectra
(continuous line) in the same
stars. For each star the top panel shows the full main beam brightness
temperature and the bottom panel corresponds to an elargement of the
same data. The epoch of the observations was October 1996 and the spectral
resolution was $0.136$ {\kms} for the $J=2 \rightarrow 1$ line and
$0.181$ {\kms} for the $J=3 \rightarrow 2$ line. To transform the
intensities into Jy one must multiply by $4.4$ Jy/K for both
transitions.}
  \label{sio32}
\end{figure*}

To compare the velocity extent of the $v=1, J=2 \rightarrow 1$
emission of SiO with the CO($2-1$) total velocity extent we define the
ratio $R_{\rm 21} = \Delta V(SiO(2-1)) / \Delta V(CO(2-1))$ where
$\Delta V$ is the full width of the emission above the $2\sigma$ noise
level at the epoch of the observations; $\Delta V(CO(2-1))$ always
refers to our 1994 observations. The values of  $R$ are given with
additional comments in
Table 2 for the 1994, 1995 and 1996 results. Because
the CO profile delineates regions of the circumstellar envelope where the
gas reaches its maximum velocity whereas the SiO
velocities trace layers close to the photosphere,
the ratio $R_{21}$ gives direct information on the kinematics of
the inner circumstellar layers. As indicated earlier, for a given star
and at each epoch,
the quantity $\Delta V(CO(2-1))$ does not vary with time in contrast with
$\Delta V(SiO)$. $\Delta V(CO(2-1))$ is
unambiguously determined although in a few cases there may be
contamination by interstellar CO (see CABG for discussion); these cases
are marked with an asterisk in Table 2. Several
stars have high $R_{21}$ values and some of them exhibit prominent red or blue
SiO wings, i.e. the SiO velocity exceeds the thermal CO velocity on the
red or blue spectral end; these cases are noted R or B in Table 2. More
than one half
of our sample shows a B or R SiO wing. The presence of red
wings in the SiO spectra indicates that there is gas emission at high
velocities not observed in the thermal CO gas. Interpretation of the SiO blue
wings is less straightforward because of possible self-absorption in the
blue wing of
the CO line profile (Huggins \& Healy 1986). Therefore, the detection of
SiO blue
wings may not necessarily mean that the CE terminal velocity is
exceeded.
If we restrict our sample of stars to Miras only, the average
value of $R_{21}$ is $\overline{R}_{21}$ = 0.70, 0.75, 0.85 and 0.86 for
the epochs 1994, 1995 and April and October 1996. The same quantities are
systematically higher for the semi-regular variables: $\overline{R}_{21}$ =
1.04,
0.85, 1.04 and 0.94. Nearly half of the observed stars have $R_{21} \geq
\overline{R}_{21}$ and 15 to 30 \% have $R_{21} \geq 1$. On the other hand,
several stars, IRC$+10011$, NML Tau, TX Cam, V Cam, GX Mon, $\chi$ Cyg and
R Cas have small $R_{21}$
values. This is perhaps due to observations made roughly at the same
stellar phase
for NML Tau, TX Cam, V Cam and GX Mon. However, some stars have low or
moderate and nearly
constant $R_{21}$ values at all phases
($\chi$ Cyg, R Cas and IRC$+10011$). We note that $\chi$ Cyg
is an S-star and therefore presents many peculiarities in the pumping
of its masers (see Bujarrabal \etal~1996). Some
others have quite high $R_{21}$ values
(e.g. T Cep and RU Her with $R_{21}=2.21$ and $2.24$ in April and October
1996, respectively).
%
\scriptsize
\begin{table*} [t]
 \hspace* {-1. cm}{ {\scriptsize{
 \caption{Total linewidth ratios at the four epochs of the observations.
 The epochs a,b,c,d correspond to Jan. 1994, June 1995, April and
 October 1996, respectively and ND means no data available.
$\Delta V(CO)$ above the $2\sigma$ noise level is given in column 2 of
this Table and refer to the 1994 observations (Cernicharo \etal 1997).
The mass loss \.{M} is taken from Loup \etal (1993).
Stars are arranged in ascending right ascension order. Comments : (*)
 CO line contaminated by galactic emission ; (R) or (B) means that the
 SiO red or blue wing reaches or surpasses respectively the most
 positive or negative CO velocity.}
 \begin{tabular}{l|c|c|cccc|cccc|ccc} \hline \hline
{\small {\bf Stars}} &{\small {\bf $\Delta V(CO)_{2\sigma}$}} & {\small
{\bf \.{M}}} & \multicolumn{4}{c}{\small {\bf Stellar
phase}}
& \multicolumn{4}{c}{\small {\bf $R_{21}=\frac{\Delta V(SiO
(2-1))}{\Delta V(CO(2-1))}$}}
& \multicolumn{3}{c}{\small {\bf $R_{32}=\frac{\Delta V(SiO
(3-2))}{\Delta V(CO(2-1))}$}} \\
  & {\footnotesize kms$^{-1}$} & {\footnotesize M$_{\odot}$/year}
&{\small {\bf a}} &{\small {\bf b}} &{\small {\bf c}} &{\small {\bf d}}
& {\small {\bf a}} &{\small {\bf b}} &{\small {\bf c}} &{\small {\bf d}}
& {\small {\bf b}} &{\small {\bf c}} &{\small {\bf d}} \\ \hline
{\bf Y Cas} & 23.6 & ND & 0.86 & 0.11 & 0.80 & 0.24
& 0.63 R & 0.65 & 0.68 & 0.72 & 0.44 & 0.20 & 0.74 R \\
{\bf T Cas} & 20.0 & 5.1~$10^{-7}$ & 0.83 & 0.004 & 0.68 & 0.10
& 0.56 & 0.99 R & 0.69 & 1.15 RB & 0.71
R & & 1.00 B  \\
{\bf R And} & 21.8 & 9.6~$10^{-7}$ & 0.15 & 0.38 & 0.13 & 0.58 &
0.93 R & 0.85 & 0.58 & 0.42 & 0.68 & & \\
{\bf IRC+10011} & 39.3 & 8.5~$10^{-6}$ & 0.65 & 0.44 & 0.90 & 0.19 &
0.30 & 0.25 & 0.36 & 0.27 & 0.21 &  & 0.30 \\
{\bf OH127.8-0.0} & 22.7 & 1.5~$10^{-6}$ & ND & ND & & & 0.21 * &
0.36 * & & & 0.45 & &  \\
{\bf $o$ Ceti} & 18.7 & 5.0~$10^{-7}$ & 0.52 & 0.94 &  & 0.67 & 0.44 &
0.85 &  & 0.80 & 0.52 & & 0.64 \\
{\bf S Per} & 35.9 & ND & 0.40 & 0.08 & 0.48 & 0.71 & 0.94 *B &
0.89 *B & 0.88 *B & 0.69 *B & 0.72 * & & 0.56  \\
{\bf NML Tau} & 40.8 & 4.4~$10^{-6}$ & 0.13 & 0.24 & & 0.29 & 0.41 &
0.32 & & 0.33 & 0.23 & & 0.39  \\
{\bf TX Cam} & 40.8 & 2.5~$10^{-6}$ & 0.33  & 0.16 & & 0.34 & 0.32 &
0.30 & & 0.46 & 0.26 & & 0.41 \\
{\bf U Ori} & 16.9 & ND & 0.22 & 0.66 & & 0.97 & 1.10 RB & 0.74
B & & 1.04 *B & 0.58 & & 0.69 *  \\
{\bf V Cam} & 31.1 & 1.6~$10^{-6}$ & 0.43 & 0.46 & & 0.39 & 0.50 &
0.59 & & 0.45 & 0.32 & & 0.34 \\
{\bf GX Mon} & 39.2 & 5.4~$10^{-6}$ & 0.40 & 0.39 & & 0.28 & 0.28 & 0.33 &
 & 0.45 & 0.32 & & 0.48 \\
{\bf S CMi} & 7.2 & ND & 0.69 & 0.14 & & 0.53 & 1.33 R & 1.60 RB &
 & 1.68 R & 0.62 R & & \\
{\bf W Cnc} & 15.7 & ND & 0.94 & 0.24 & & 0.50 & 0.92 R & 0.69 &
 & 0.78 R & 0.18 & & \\
{\bf R LMi} & 19.5 & 2.8~$10^{-7}$ & 0.34 & 0.70 & & 0.16 & 0.70 R &
0.81 R & & 0.85 R & 0.56 & & \\
{\bf R Leo} & 15.7 & 1.0~$10^{-7}$ & 0.63 & 0.29 & & 0.85 & 1.00 RB & 1.00
R & & 1.20 R & 0.87 R & & 0.91 R \\
{\bf R Crt} & 23.5 & 7.5~$10^{-7}$ & ND & ND & & ND & 0.82 B & 1.07
B & & 0.92 & 0.68 & & \\
{\bf RT Vir} & 18.8 & 7.4~$10^{-7}$ & 0.95 & 0.58 & 0.56 & 0.77 &
1.22 B & 1.06 B & 1.14 B & 0.87 & 0.58 & & 0.60  \\
{\bf SW Vir} & 14.9 & 5.7~$10^{-7}$ & 0.69 & 0.05 & 0.98 & 0.27 &
0.55 & 0.75  & 1.22 B & 1.13 B & & & \\
{\bf R Hya} & 15.9 & 1.4~$10^{-7}$ & 0.92 & 0.30 & 0.00 & 0.57 & 0.53 &
0.96 R & 0.95 R & 0.85 & 0.90 R & & 0.40 \\
{\bf W Hya} & 17.2 & 8.1~$10^{-8}$ & 0.68 & 0.01 & 0.92 & 0.34 &
1.30 RB & 0.97 B & 1.00 RB & 0.97 B &1.06 R & & \\
{\bf RX Boo} & 19.5 & 8.1~$10^{-7}$ & 0.81 & 0.21 & 0.99 & 0.67 &
0.65 & 1.06 B & 1.05 RB & 1.03 *RB & 0.41 & 0.46 & 0.30 * \\
{\bf S Crb} & 15.0 & 5.8~$10^{-7}$ & 0.26 & 0.74 & 0.51 & 0.03 & 0.95
R & 0.77 & 1.00 B & 1.17 B & 0.34 & & 0.69 \\
{\bf RU Her} & 16.4 & ND & 0.10 & 0.17 & 0.81 & 0.19 & 0.85 R & 0.92 B &
0.93 B & 2.24 R & 0.85 & 0.34 & 0.67 \\
{\bf U Her} & 19.9 & 2.6~$10^{-7}$ & 0.66 & 0.03 & 0.77 & 0.23 &
0.74 & 0.82 & 0.89 R & 0.92 & 0.72 & 0.61 & 0.97 R \\
{\bf R Aql} & 18.8 & 3.6~$10^{-7}$ & 0.004 & 0.02 & 0.03 & 0.69 &
0.62 & 0.85 B & 0.79 R & 0.57 * & 0.43 & & 0.38 *  \\
{\bf $\chi$ Cyg} & 24.8 & 5.6~$10^{-7}$ & 0.70 & 0.99 & 0.75 & 0.21 &
0.50 & 0.66 & 0.53 & 0.60 & 0.73 & & 0.52 \\
{\bf T Cep} & 11.8 & 1.4~$10^{-7}$ & 0.04 & 0.35 & 0.13 & 0.61 &
1.07 RB & 1.36 RB & 2.21 RB & 1.49 RB & 1.02 B & & 1.19 R \\
{\bf $\mu$ Cep} & 35.9 & ND & 0.81 & 0.43 &  & 0.65 & 1.83 *RB &
0.50 * & & 0.84 *B & 0.51 * & & 0.56 * \\
{\bf R Cas} & 28.8 & 1.1~$10^{-6}$ & 0.35 & 0.60 & 0.28 & 0.71 &
0.66 & 0.52 & 0.59 & 0.38 * & 0.52 & 0.50 & 0.23 * \\ \hline \hline
\end{tabular}}}}
 \label{tablephaseR}
\end{table*}
\normalsize
\subsection{SiO $v=1, J=3 \rightarrow 2$ emission}

 The central velocity of the $v=1, J=3 \rightarrow 2$ emission is
often quite different from that of the $J=2 \rightarrow 1$ emission
as one would expect in gas layers with different excitation
conditions. With a few exceptions, the line intensities and
full linewidths above the $2\sigma$ noise level for the $J=3\rightarrow
2$ line are systematically smaller than for the $J=2 \rightarrow 1$ line.
As an example, in June 1995 we have
obtained for the Miras R Leo and U Her, $T_{\rm MB}$ = 19.4
and 8.1 K, $\Delta V$ = 13.7 and 14.3 {\kms} in the $J=3 \rightarrow 2$
line while we have measured $T_{\rm MB}$ = 50.2
and 61.3 K, $\Delta V$ = 15.8 and 16.4 {\kms} in the $J=2 \rightarrow
1$ line. Examples of spectra are given in Fig. \ref{sio32}. As for
the $J=2 \rightarrow 1$ line we define the ratio
$R_{\rm 32} = \Delta V(SiO(3-2)) / \Delta V(CO(2-1))$, and we give
the individual values of this ratio in the last three columns of
Table 3.
If we restrict our sample of stars to Miras only, the average
value of $R_{32}$ is $\overline{R}_{32}$ = 0.57, 0.41 and 0.61 for
the epochs 1995, and April and October 1996. In contrast with
$\overline{R}_{21}$, the values of $\overline{R}_{32}$ are not
systematically higher for the semi-regular variables: $\overline{R}_{32}$ =
0.55,
0.68 and 0.52 for the epochs 1995, and April and October 1996.
All values of $R_{32}$ and $\overline{R}_{32}$ are lower than the
corresponding values derived for the $J=2
\rightarrow 1$ line; this reflects the fact that $\Delta V(SiO(3-2)) <
\Delta V(SiO(2-1))$. Nevertheless, in June 1995 and October 1996 where 20
to 30 stars where observed at 2 mm, few of them have a
value of $R_{32}$ larger than, or of order 1. The case of T Cep is
interesting as the large $R_{21}$ value observed at 3 mm
is also observed at 2 mm (Fig. \ref{sio32} and Table 2). About 20\%
of the observed stars show a blue or red SiO wing with a majority of
stars exhibiting a red wing (Table 2).

\subsection{Variability of the ratio $R=\Delta v(SiO) / \Delta v(CO)$}
\label{variability}
\begin{figure*} [t]
  \begin{center}
     \epsfxsize=15cm
     \epsfbox{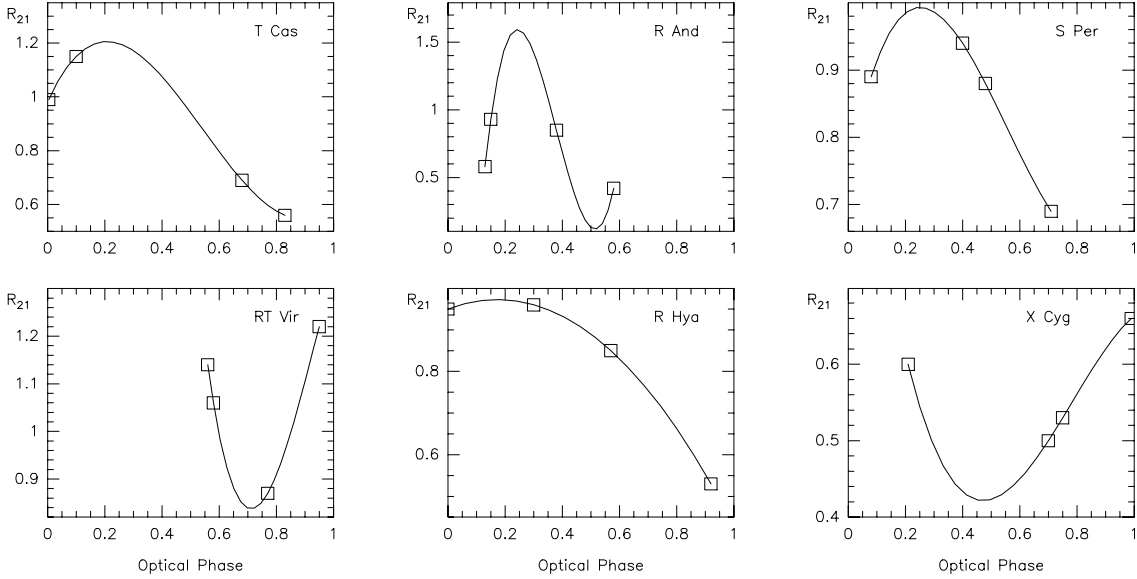}
  \end{center}
  \caption []{Examples of variations of $R_{\rm 21} = \Delta V(SiO(2-1)) /
\Delta
V(CO(2-1))$ with the stellar phase for the semi-regular variables S Per
and RT Vir, and the Miras R And, R Hya, $\chi$ Cyg and
T Cas. The linewidths are defined at the $2\sigma$ noise level.}
  \label{R_phase}
\end{figure*}

 In Table 2 we compare the values of $R_{\rm 21}$
derived for $\Delta V(SiO(2-1))$ at the four
epochs of the observations and we derive the associated stellar phase
from the most recent stellar light parameters given by the
AFOEV\footnote{AFOEV=Association Francaise des Observateurs
d'Etoiles Variables}. The stellar
period was rather well sampled in general although
additional observations would be deserved. Our data
show that there is a dependence of $R_{21}$ with the optical phase
for each star. This dependence is clearly not similar for all stars in
our sample. Despite the fact that we only have 3 or 4
distinct epochs and that the stellar periods are unequally sampled we have
searched for
periodicity in our data by fitting cubic spline functions.
We show six examples in Fig. \ref{R_phase}.
\begin{figure} [t]
  \begin{center}
     \epsfxsize=6cm
     \epsfbox{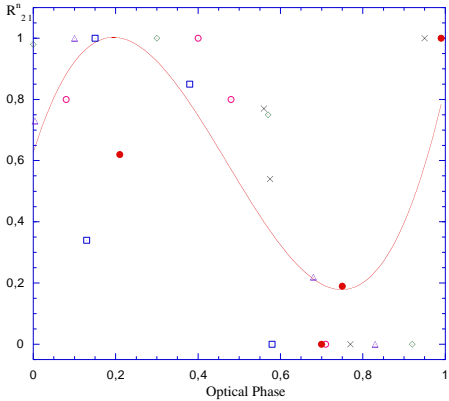}
  \end{center}
  \caption []{Plot of the normalized ratio ${R_{\rm 21}}^{n} = \frac{\Delta
  V(SiO(2-1))} {\Delta
V(CO(2-1))}$ versus the stellar phase for the stars plotted in
Fig.\ref{R_phase}. Each type of plot marker corresponds to a different
star: $\bullet =$ $\chi$ Cyg, $\bigtriangleup =$T Cas,
$\Box =$R And, $\times =$RT Vir, $\diamond =$R Hya, $\circ =$S Per.
The fit is a third degree polynomial.}
  \label{rcorrphase}
\end{figure}
In several cases $R_{\rm 21}$ reaches a maximum and a
minimum around the optical phases $0.1-0.3$ and $0.6-0.8$, respectively.
This maximum activity seems to occur with the same phase
lag as that observed for the bulk of the SiO emission ($0.2-0.3$ in
general). The general trend
above is not always observed. In R Cas or $\chi$ Cyg for example, $R_{21}$
tends to be maximum around 0.0, and in the case of
W Hya (SRa) $R_{\rm 21}$ tends to remain roughly
constant with time.
As the amplitudes of the variations of
$R_{21}$ are obviously different from one star to
the other, and because our sample of stars is rather homogeneous,
we have normalized these values by forcing the maximum and minimum
$R_{21}$ values for each star to 1.0 and 0.0, respectively; these normalized
values, ${R_{21}}^{n}$, are plotted versus the
optical phase in Fig.\ref{rcorrphase}. A global trend is obvious with a slight
maximum and minimum of ${R_{21}}^{n}$ between optical phases 0.1-0.3 and
0.6-0.8, respectively, as for
the individual plots.
\begin{figure} [ht]
  \begin{center}
     \epsfxsize=8cm
     \epsfbox{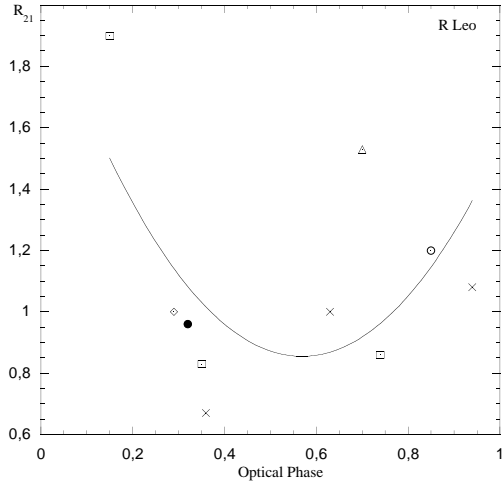}
  \end{center}
  \caption []{Plot of the ratio $R_{\rm 21} = \Delta V(SiO(2-1)) / \Delta
V(CO(2-1))$ versus the stellar phase for R Leo.
The linewidths are defined at the $2\sigma$ noise
level. Each type of plot marker corresponds to a different stellar
cycle: $\bullet =$April 1991, $\bigtriangleup =$June 1992,
$\Box =$Nov. 1992 to May 1993,
$\times =$Nov. 1993 to May 1994, $\diamond =$June 1995, $\circ =$October 1996.}
  \label{rleo}
\end{figure}

In some stars, R And, TX Cam, R Leo, RU Her and R Aql,
we have observed a rapid
variation in the amplitude of $R_{\rm 21}$. These variations occur nearly
at the same stellar
phase but correspond to
observations made in 1994, 1995 and 1996, and thus correspond to
different stellar pulsation cycles. This suggests
that besides the smooth variation of $R_{21}$ with the stellar
pulsation observed in several stars, the SiO wing emission
may differ from one optical cycle to the
other. We may thus have the combined effect of a smooth stellar pulsation with
another mechanism (of a different time order). This remark also
applies to R Leo (Fig. \ref{rleo}) for which several epochs are
available although on average $R_{21}$ tends to be maximum and
minimum around optical phases $0.2$ and $0.5$, repectively.
Comparing the behaviours of $R_{21}$ and of the stellar visual
magnitude $m_{v}$ with time, we note that the sudden amplitude variations
of $R_{21}$ at a given optical phase (as observed for example in R Aql and
R Leo
near optical phases 0.0 and 0.7 respectively), occur simultaneously
with strong amplitude variations of $m_{v}$ from one stellar cycle to
the other. In all cases, the sudden increase in $R_{21}$ corresponds
to a luminosity increase which one may relate to IR pumping of these masers.

In Table 2 we also give the ratio
$R_{\rm 32}$ and the optical phase at the four epochs of our
observations. The general trends observed for
$R_{\rm 21}$ are also present in $R_{\rm 32}$. However, the SiO line
wing intensities and extents are smaller in the  $J=3 \rightarrow 2$ line than
in the $J=2 \rightarrow 1$ line. RU Her is an extreme case where the
conspicuous red wing detected in October 1996
at 3 mm is not present in the 2 mm spectrum.

\begin{figure} [ht]
  \begin{center}
     \epsfxsize=8cm
     \epsfbox{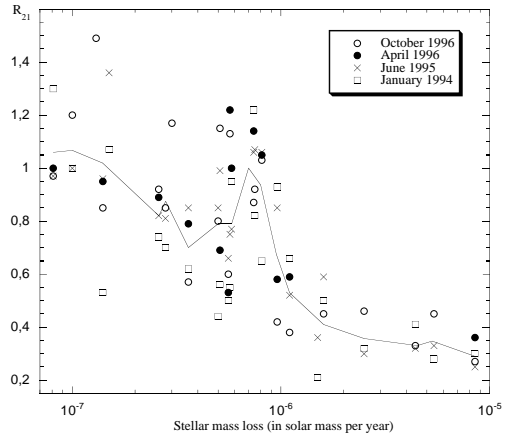}
  \end{center}
  \caption []{Plot of $R_{\rm 21}=\Delta V(SiO(2-1)) / \Delta
V(CO(2-1))$  versus the stellar mass loss (in M$_{\odot}$/year) for
all observations and all stars. The mass loss is taken from Loup
\etal (1993). The solid line connects the averages of 
all points as determined at individual stellar mass loss values.}
  \label{pertetotal}
\end{figure}
 It is interesting to compare the time variation of the ratio
 $R_{21}$ with that of the SiO peak
intensity which is associated with the bulk of the emission
(characterized by $T_{MB}$ and the associated central velocity).
The SiO main emission is known to vary from one stellar
cycle to the other and to reach a maximum somewhat after the optical
maximum (e.g. Alcolea 1993, and Martinez et al. 1988). As previously
observed by
other authors, we also find that $T_{\rm MB}(SiO(2-1))$ varies with the
optical phase.
However, because the stellar cycle is not densely enough sampled it is
impossible to conclude that $R_{21}$ and $T_{MB}$ vary in phase or
not, even if $R_{21}$ seems to present a maximum and minimum at the
same phases as usually observed for the intensity.

\section{Discussion}
\label{discussion}

 In this section we first discuss some of the mechanisms at the origin
of the high velocity SiO emission and whether these are consistent
with the observed behaviour of the ratio $R$ with the light cycle.
Secondly, we will discuss whether our data bring some evidence or not
on the question of the co-location of the SiO
wing emission and of the bulk of the SiO emission, and on the
influence of the stellar mass loss rate.

\subsection{Formation of SiO high velocity wings}

 Various mechanisms, or a combination of these mechanisms, may be
invoked to explain the high velocities observed in our SiO spectra:
turbulent motions, stellar pulsation, rotation or asymmetric mass
loss. These mechanisms were discussed earlier by CABG, mainly they
demonstrated the importance of the asymmetric mass loss while the
impact of rotation was shown to be minor. In addition, the
characteristic profile predicted by rotation (e.g. two- or
three-horn profile) is not observed in our
SiO spectra. Our new data
bring information on the two first mechanisms, namely turbulent
motions and stellar pulsation.

\subsubsection{Turbulent motions}

 With the non-local radiative transfer code described by
Gonz\'{a}lez-Alfonso \& Cernicharo
(1997), CABG modeled the SiO emission in an
expanding envelope with low terminal velocity, high mass loss rate and
strong turbulent motions. CABG succeeded in predicting blue
wings and weak red wings because of shadowing and amplification of
stellar emission. Several
stars observed in this work show indeed blue wings
without a red counterpart at a given optical phase. This is observed
for 7 of the 8 semi-regulars in our sample and for 4 Miras. In the 4
semi-regulars S Per (SRc), R Crt (SRb), RT Vir (SRb) and SW Vir (SRb)
we have not detected any red wing at any epoch. Turbulence could thus
play a role in the formation of blue wings in semi-regular variables
and rarely in Miras. However, we did not observe that all
semi-regulars and Miras showed blue wing emission without red
counterpart whichever optical phase we observed. This tends to prove that
turbulence does not dominate, but may nevertheless contribute to the
formation of
the line wings.

\subsubsection{Stellar Pulsation}

 The fact that our data show for most stars in our sample a rather
regular pattern for the variations of $R_{21}$ with the optical phase
demonstrates that some connection exists between the stellar
pulsation and the occurence of SiO linewings. Therefore, the stellar pulsation
induces variations of the ratio $R_{21}$ with a maximum activity consistent
with
the 0.1-0.3 phase lag observed in the bulk of the SiO emission with
respect to the stellar flux. (In fact, the SiO maser main intensity
follows the IR luminosity with optical phase lags $0.1-0.3$ and
$0.6-0.8$ for maxima and minima, respectively.) Several observations in R
Leo confirm
this result. But the stellar pulsation can induce variations in the line
wing emission in two different ways: directly by shocks, or
indirectly by luminosity variations. Concerning the shocks, several models
have
attempted to predict the behaviour of the outer atmospheric gas layers in
pulsating
late-type variables (e.g. Bowen 1988, Wood 1989). The non-linear
pulsation model of Wood (1989) successfully explains the velocity
changes observed for the hydrogen lines in Miras. The shock wave
generated at each pulsation cycle is responsible for the velocity
discontinuity. This is also in agreement with the IR line observations
of Hinkle \etal (1984) showing both photospheric material and
outwardly moving gas (post-shock material), and
with the results of Boboltz \etal (1997) who observed SiO masers undergoing
infall in R Aqr.
In these conditions the
radial velocities may exceed the terminal velocity of the associated
circumstellar envelope as this is the case for our SiO observations.
Similar predictions are made by Bowen's (1988) model where the stellar
pulsation generates outflowing and infalling gas layers which could be
responsible for our blue and red SiO wings. Velocities larger than the
terminal velocity are also predicted in the innermost layers of the
envelope. In fact, the outflowing gas layer running into the
infalling gas creates a shock wave which locally modifies the physical
conditions, and in particular increases the local density of hydrogen,
making maser
emission more favourable at these higher velocities. (Note that this
does not necessarily imply that collisions are dominant although the
results of Miyoshi \etal (1994) suggest that
collisional pumping plays a role for the bulk of the SiO emission.) On the
other hand,
if radiative pumping dominates, the observed variations of the line wing
emission should correspond to maximum pumping efficiency around
the optical phase 0.2
and minimum efficiency around phase 0.6. A change by a small factor in the
fully
saturated peak emission may imply an enormous factor for the unsaturated
wings.
In conclusion stellar flux variations and shock waves are
important to explain our observations and maser emission, even if it is
not easy to trace shocks by studying the SiO maser. Shocks may
explain that there is SiO emission at high velocities, and radiative
pumping may explain the variations of $R_{21}$.

The strong variations at a given phase from one cycle to
another of $R_{21}$, of the main emission
temperature $T_{MB}$ and of the velocity observed in R And, TX Cam, RU Her
and R Aql, as well as the changes in the SiO linewing emission observed in
R Leo for different stellar cycles suggest that
variations of the stellar physical conditions imply sudden changes in the
SiO wing and main line emissions. These variations cannot be ascribed to
changes
in $\Delta V(CO)$ which was found to remain constant at all epochs.
Sudden changes are perhaps related to non-periodical stellar structure
changes implying changes in the SiO masing conditions.
On the other hand, we cannot exclude the hypothesis of an overtone
stellar pulsation mode which could induce short-time variations
of the physical parameters, and thus of the SiO emission.
Rapid changes of the main and wing line emissions could
also be explained by sound waves which
induce local variations in the density and relative velocity just above
the stellar photosphere. We thus expect SiO line variations within short
time scales. Pijpers \etal (1994) observed indeed in R Leo and R Cas
variations of the intensity and velocity of the $v=1, J=1\rightarrow 0$
line over short periods of order 10-20 days. These sound
waves could modify the SiO main and wing line emissions if both
emissions were co-located.

\subsection{Location of the SiO line wings}

Interferometric observations of the SiO line wings would of course be
indispensable to accurately locate the red and blue line wing emissions
within the circumstellar envelope. There is at least one known case,
that of R Leo, for which blue and red SiO wings are not coexistent. This
was demonstrated by the lunar occultations made by Cernicharo \etal (1994)
and by the relative position measurements made with the IRAM
interferometer (Baudry \etal 1995). We note that, in the absence of
interferometric data, the observation of phase lags between the SiO
wing and bulk emissions could in principle indicate different spatial
locations. However, in the case of
variations in the radiative pumping resulting from stellar flux
variations one would expect very small and thus not observable phase lags,
whereas in the case of dominant shock fronts the propagation
is slower, and may induce only very long-term phase lags. On the other
hand, and despite our lack of more densely sampled observations,
our study of the ratio $R_{21}$ with the stellar light cycle
gives indirect evidence for similar conditions of excitation for
the SiO wings and the bulk of the SiO emission. We
have found that the
maximum value of the ratio $R_{21}$, namely the maximum activity in
the line wings, tends to occur when the maximum activity also occurs
in the bulk of the SiO emission. This seems to be the general trend
for most stars in our sample (see Sect. \ref{variability}). In addition,
for those stars in which we have observed sudden variations of the
amplitude of $R_{21}$ we have observed simultaneous variations of
$T_{MB}$ and velocity. This suggests again
that the main SiO line emission and the SiO line wings require similar
pumping mechanisms, although they might not be excited in the same
gas layers with similar physical conditions (e.g. different SiO
column densities).

In order to deepen the physical meaning of the ratio $R$, we
have plotted the values of $R$ obtained at the four epochs of our
observations versus the stellar mass loss rate $dM/dt$ taken from Loup et al.
(1993). For both $R_{\rm 21}$ and $R_{\rm 32}$,
$R$ decreases with increasing mass loss rate, with two notable
features for $R_{21}$ (Fig. \ref{pertetotal}): a sudden
increase for mass loss rates around $5-9~10^{-7}$
M$_{\odot}$/yr, and a broad plateau where $R_{21}$ stays low and
roughly constant for $dM/dt$ between
$10^{-6}$ and $10^{-5}$ M$_{\odot}$/yr. We note, however, that the main
features
in Fig. \ref{pertetotal} would deserve confirmation with additional
data, and that the scatter present in the plot $R_{21}=f(dM/dt)$ may
also be due to uncertainties in the mass loss rates. In fact, our
sample is biased because of a dependence of $dM/dt$ on the
distance in the range 100 pc to 400 pc (sensitivity limitation).
Nevertheless, even for the most
distant stars, the detection rate of line wing emission is comparable to
that for the closest stars. Because $\Delta V(CO)$ is a good indicator
of the expansion of the envelope while $\Delta V(SiO)$ is closely
related to both the excitation and the kinematics of SiO we have
plotted separately $\Delta V(CO)$ and $\Delta V(SiO)$ versus $dM/dt$
for our sample stars. As expected for a species excited far in the
envelope, $\Delta V(CO)$ increases with $dM/dt$; we also observe a
flattening for the highest mass loss rates. On the other hand,
$\Delta V(SiO)$ does not clearly exhibit the general decrease
observed in the plot $R_{21}=f(dM/dt)$. Therefore, this decrease
as well as the flattening observed for $R_{21}$ seem
to be readily explained by the general behaviour of $\Delta V(CO)$ in
our sample. These plots demonstrate that the regions of SiO wing
emission and CO expansion are not directly related and strongly
support the idea that the SiO wing emission arises from the innermost
circumstellar layers (although we cannot say it coexists entirely
with the bulk of the SiO emission). The apparent peak of SiO wing
emission observed in Fig. \ref{pertetotal} suggests that, to excite
this emission, more favourable pumping conditions (e.g. adequate
densities) may exist around $5-9~10^{-7}$ M$_{\odot}$/yr. However, we
are unable at the moment to specify error bars on the mass-loss rates,
and therefore the reality of the peak for the ratio $R_{21}$ must be
confirmed by newer observations obtained for a larger sample of stars.

\section{Conclusions}

 We have observed a rather large sample of late-type stars including
Miras and semi-regulars
with the IRAM 30-m radiotelescope at four epochs covering the period 1994
to 1996 in order to investigate the correlation between the SiO linewing
activity and
the stellar light phase. The SiO $v=1, J=2 \rightarrow 1$ and $J=3
\rightarrow 2$
lines were observed simultaneously with the CO $J=2 \rightarrow 1$
quasi-thermal emission line.
Several high velocity wings have been detected in the red and blue
edges of the SiO profile.

The SiO wing emission could result from complex mechanisms combining
stellar pulsation of the fundamental
mode, and perhaps of other modes, with asymmetric mass loss (as for R
Leo) and
structure changes. We deduce from our observations that the time evolution of
the SiO line wings is related to the stellar
pulsation. However, it is difficult
to specify how the pulsation induces local physical variations
responsible for variations of the SiO wing emission. Pulsation,
through shocks, may produce high velocity emissions which then vary
according to the optical phase. For the semi-regular variables there
is some indication on the importance of turbulent motions in
the formation of the high velocity emission as well.

Finally, we have shown that the SiO wing emission results from masing
processes and that this emission very likely arises from the
innermost gas layers of the circumstellar envelope.

\acknowledgements
{We thank the IRAM personnel at the 30-m telescope for their
efficient help during the observations and we thank P.J. Diamond and E.
Gonz\'alez-Alfonso
for very useful comments. Part of this work was supported by the CNRS URA 352.
This research has made use of the AFOEV database, operated at CDS, France.}

\end{document}